\documentclass{article}

\newif\ifpaper
\papertrue

\usepackage{cp_shared/arxiv}
\usepackage[style=authoryear,doi=true,isbn=true,url=false,eprint=true]{biblatex}

\usepackage[utf8]{inputenc} 
\usepackage[T1]{fontenc}    
\usepackage{hyperref}       
\usepackage{url}            
\usepackage{booktabs}       
\usepackage{amsfonts}       
\usepackage{nicefrac}       
\usepackage{microtype}      
\usepackage{graphicx}

\usepackage{doi}

\usepackage{gensymb} 
\usepackage{rotating} 
\usepackage{caption}
\usepackage{subcaption}
\usepackage[nobottomtitles*]{titlesec}
\usepackage{amsmath}
\usepackage{setspace}

\newcommand{\code}[1]{\lstinline[language=bash, basicstyle=\ttfamily\small]|#1|}
\usepackage{color}
\definecolor{lightgrey}{rgb}{0.975, 0.975, 0.975}
\definecolor{midgrey}{rgb}{0.6, 0.6, 0.6}
\definecolor{deepgrey}{rgb}{0.2, 0.2, 0.2}
\definecolor{codegreen}{rgb}{0, 0.7, 0}
\definecolor{codepink}{rgb}{0.8196, 0.10, 0.654}
\usepackage{listings}
\title{The application of mixed-use measures at the pedestrian-scale}

\date{} 

\author{
	\href{https://orcid.org/0000-0003-3790-0638}{
		\includegraphics[width=0.25cm, height=0.25cm, keepaspectratio]{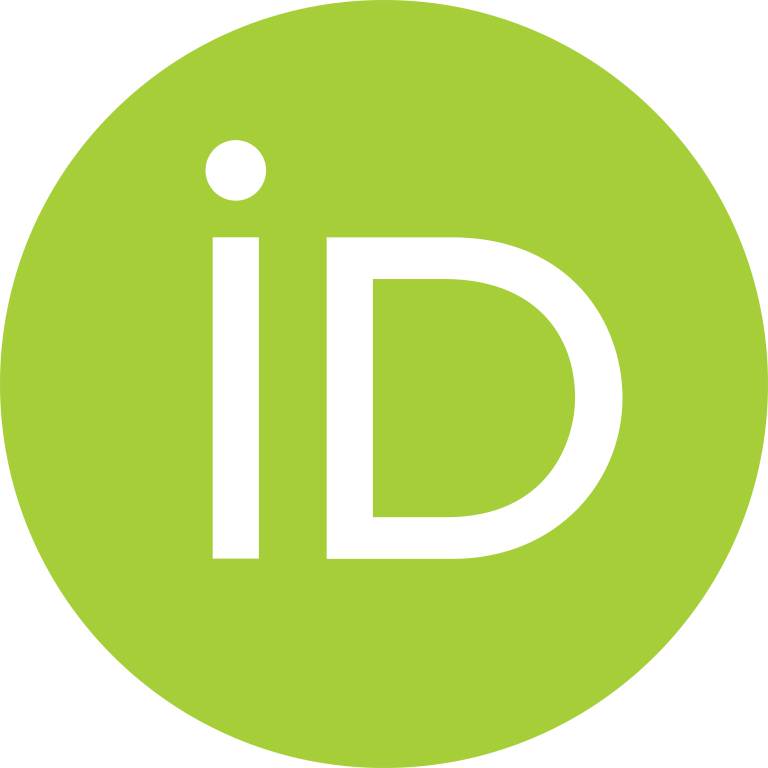}
		\hspace{1mm}Gareth D. Simons
	}
	\thanks{Benchmark Urbanism \texttt{gareth@benchmarkurbanism.com}}
}

\renewcommand{\shorttitle}{Mixed-uses at the pedestrian-scale}

\hypersetup{
	pdftitle={
		The application of mixed-use measures at the pedestrian-scale
	},
	pdfsubject={
		physics.soc-ph
	},
	pdfauthor={
		Gareth~D.~Simons,
	},
	pdfkeywords={
		computation,
		data science,
		land-use analysis,
		morphometrics,
		network analysis,
		spatial analysis,
		urban analytics,
		urban planning,
		urban morphology,
		urbanism
	},
}

\begin{document}
\maketitle
\begin{abstract}
	Mixed-use urbanism affords access to diverse assortments of land-uses within a pedestrian-accessible context. It confers advantages such as reductions to driving, air pollution, and Body Mass Index with associated increases in active transportation and improvements to health. However, whereas mixed-use urbanism is clearly beneficial, methods for measuring and assessing the presence of mixed-uses at a granular level of analysis remain murkier.

This work demonstrates techniques for gauging mixed-uses in more spatially precise terms concurring more readily with an urbanist's conception of pedestrian-accessible mixed-uses. It does so through the use of the \code{cityseer-api} \code{Python} package, which facilitates the use of spatially granular land-use classification data assigned to adjacent street edges and then aggregated dynamically, with distances measured from each point of analysis to each accessible land-use while taking the direction of approach into account. It is argued that \emph{Hill Numbers} is a suitable measure of diversity because it can mirror the intent of traditional indices while behaving more intuitively. Further, distance-weighted formulations of Hill diversity can be applied with spatial impedances, thus conferring a particularly spatially nuanced gauge of local access to mixed-uses.

These methods and indices are demonstrated for Greater London with observations correlated to Principal Component Analysis derived from a range of land-use accessibilities measured from the same locations and for the same point-of-interest dataset. The Hill diversity measures, particularly the distance-weighted formulations, offer the most robust correlations for both expansive mixed-use districts and more local `high-street' mixes of uses while yielding the most intuitive and spatially precise behaviour in the accompanying plots.
\end{abstract}
\section{Mixed-uses, modernity, and suburbia}\label{mixed-uses-modernity-suburbia}

The agglomeration of people is, necessarily, underpinned by a substantial complexity of interactions, and this is reflected in the variety and propinquity of land-uses afforded by urbanisation \parencite{Glaeser1992, Florida2003}. The evolution of mobility and communications technologies has dramatically intensified and fragmented these dynamics across space, yet has not nullified the beneficial aspect of geographic proximity nor the human desire for contact and a sense of place \parencite{Graham2001, Jeffres2002}. Within the context of urbanism, the term \emph{mixed-uses} refers to assortments of diverse land-uses facilitating varied assortments of interaction, but with the specific connotation that these should be available to pedestrians at the local scale. In effect, this is a pedestrian-first rather than car-centric premise for urban connectivity, not to be construed as nostalgia for traditional neighbourhoods or idealised small-town living. Note that not all land-use locations necessarily need to be within strict walking distances if, as may be typical for larger towns and cities, these are accessible in concert with transit or active transportation.

The interconnection of land-uses has always been desirable and prevalent throughout the history of urbanisation. However, a thorny issue for planners has been that revolutionary forms of personal motorised transportation and the emergence of communications technologies led to the assumption that it was possible, even preferable, to sustain these interactions purely at the larger scale and that it was acceptable to do so at the expense of local pedestrian connectivity. First triggered by the emergence of commuter railways, early instances of land-use separation promised an escape from the squalor of industrial cities to the `slumless' and `smokeless' satellite suburbs idealised by the \emph{Garden City Movement}. Though finding support in early modernist planning philosophies \parencite{Garnier1989, Corbusier1967} it would be the mass-production of cars that would ultimately fuel the increasingly blatant and indiscriminate separation of land-uses, ultimately finding its most pathological expression in post World War II American suburbia \parencite{Fishman1987}.

The history of suburbia veers into a broader discussion of modernity and the destabilisation of traditional identities, a pervasive search for a sense of place and belonging, and the collective pursuit of private interests. Early 20th century sociologists documented a momentous societal shift from a predominately rural to urban state of existence in western societies. Traditional \emph{folk society}, typified by small and familiar, often isolated, and culturally homogenous towns and villages with a strong sense of solidarity \parencite{Redfield1947} gave way to an increasingly urbanised existence characterised by greater social anonymity and exposure to heterogenous assortments of cultures and ideas. Cities swelled as rural inhabitants bearing the imprint of a rural past were recruited from the countryside; yet, though typically envisioned in one direction, an opposite dynamic was also at work: rural dwellers were increasingly affected by forms of culture and technology emanating from cities \parencite{Wirth1938}. The relentless development of mobility and communication technologies expanded spheres of social and economic interaction with the consequent intensification of interactions shifting relationships away from the realm of the familiar into the desensitised state of the blas\'e: impersonal, segmented, and superficial with qualitative distinctions increasingly reduced to monetary terms \parencite{Simmel1997}. Rapid urbanisation thus presented a paradox: it was at once both liberating and deeply unsettling. Traditional communities had offered a strong --- albeit prescriptive --- sense of identity and a more stable --- albeit geographically confined --- sense of place and belonging. Cities, in contrast, implied a more fluid and competitive dynamic emerging from an unstable equilibrium maintained through mobility and continual adaptation. Identity now had to be constructed from transitory assemblages of ideas, people, and places \parencite{Park1915, Lyon1999}, provoking a collective sense of loss and nostalgia for a past life idealised as more straightforward, geographically anchored, and socially homogenous, beacons of meaning and security in a world beset by constant change \parencite{Jeffres2002, Ellin1999}.

A dilemma confronted city dwellers: repulsed by the estrangement and perceived evils of urban life, they were, nonetheless, attracted to the opportunities afforded by urbanisation. Suburbs are an attempt to synthesise this dichotomy through the collective pursuit of individual interests behind the facades of harmony with nature and close-knit stable communities \parencite{Fishman1987}. Mass-production of cars coupled with access to cheap land unleashed the leap-frogging suburban development that has become a ubiquitous template: sprawling and fragmented assemblages of low-density, single-land-use zones patch-worked together by motorways. Though current planning policies may give a different impression, these patterns remain tangible and feature in contemporary forms of urban development worldwide. By way of example, a 2018 report based on a qualitative review of new housing developments in the United Kingdom found car-based living; homes not properly connected for pedestrians, cyclists, or busses; missed opportunities for public transport; and a lack of mixed-uses \parencite{TransportforNewHomes2018}.

A twist of irony remains. The suburban ideal imploded into \emph{suburbia}, an unanchored and endlessly repeating stream of motorways, malls, and fast-food signs, exuding an all-pervasive sense of placelessness that only exacerbated the sense of loss predicated by modernity \parencite{Lyon1999, Ellin1999}. A consequent backlash against suburban planning emerged in the form of \emph{New Urbanism}, a set of planning principles based on traditional neighbourhoods prevalent before the widespread emergence of suburbia, broadly advocating the use of public spaces, mixed land-uses, pedestrian-friendly design, and the clear articulation of public and private spatial thresholds \parencite{Katz1994, Langdon1994, Kunstler1996, Duany2000, Calthorpe1993}. Whereas these design principles are generally sound, New Urbanism also echos the now-familiar refrain: the desire for a sense of place and community associated with the past and the notion that replicating traditional neighbourhood design principles leads to more robust local communities. As such, New Urbanism has been prone to pastiche due to a tendency to conflate the concept of \emph{community}, in the broader sociological sense of the word, with that of the \emph{neighbourhood}, a smaller subset of geographically anchored social interactions \parencite{Scully1996, Jeffres2002}. Whereas traditional neighbourhood design principles may encourage casual and neighbourly social interaction \parencite{Skjaeveland1997, Haggerty1982}, these do not necessarily lead to deeper social bonds solely based on propinquity \parencite{Talen1999, Handy1992, Lund2002, Audirac1994, LundHollie2003}.

Quantitative support for mixed-use development is emerging around themes on reductions in driving, air pollution, and Body Mass Index, as well as increases in the use of active transportation and improvements to health \parencite{Saelens2003, Brown2009, Frank2006, DeNazelle2011, Sallis2016}. Potential benefits to local economic vibrancy and the resilience of neighbourhoods, as advanced by \textcite{Jacobs1961}, tend to remain less developed and present a rich avenue for exploration. Unlike road networks, which evolve slowly and rarely change, granular mixed-use districts adapt comparatively quickly in response to evolving economic opportunities and constraints: a reflection of underlying complex-adaptive urban systems and their capacity for resilience through feedback and adaptation --- otherwise incapacitated by the crude brush-strokes of abstract urban planning \parencite{Batty2012, Marshall2009, Marshall2012}.
\section{Land-use classification schemas and taxonomic limits}\label{land-use-classification-schemas-and-taxonomic-limits}

This section introduces the \emph{Ordnance Survey} \emph{Points of Interest} dataset through a cursory review of the relationship between land-uses and city sizes. Various properties of towns and cities scale with an increase in population size, and several theories have been proposed to describe the possible mechanisms behind such allometric transformations \parencite[see][p.38]{Batty2013}. Of interest to mixed-uses is the total quantity and uniqueness of land-use classifications in relation to city size. The hypothesis is that increases in population size prompt increased social and economic opportunities and land-use diversity. This concept bears a resemblance to species-area laws from ecology, which hold that more significant landmasses have more resources and a greater variety of habitats at their disposal, thus supporting larger and more diverse assortments of species. This theory was demonstrated by \textcite{Macarthur1963} for islands by intersecting species immigration and extinction curves and was subsequently extended to a more generalised species-energy law \parencite{Wright1983}. The relationship tends to be described in the power-law form $S=cA^z$, where the number of unique species $S$ is a function of the constant $c$ and the land-area $A$ raised to the exponent $z$, which typically assumes a value in the vicinity of $z \approx 0.25$ \parencite{Gould1979}. This relationship has, historically, also been ascribed to a semilog form of $S=c+zlog(A)$ \parencite{Gleason1922}.

\begin{figure}[htbp]
 \centering
 \includegraphics[width=\textwidth,height=\textheight,keepaspectratio]{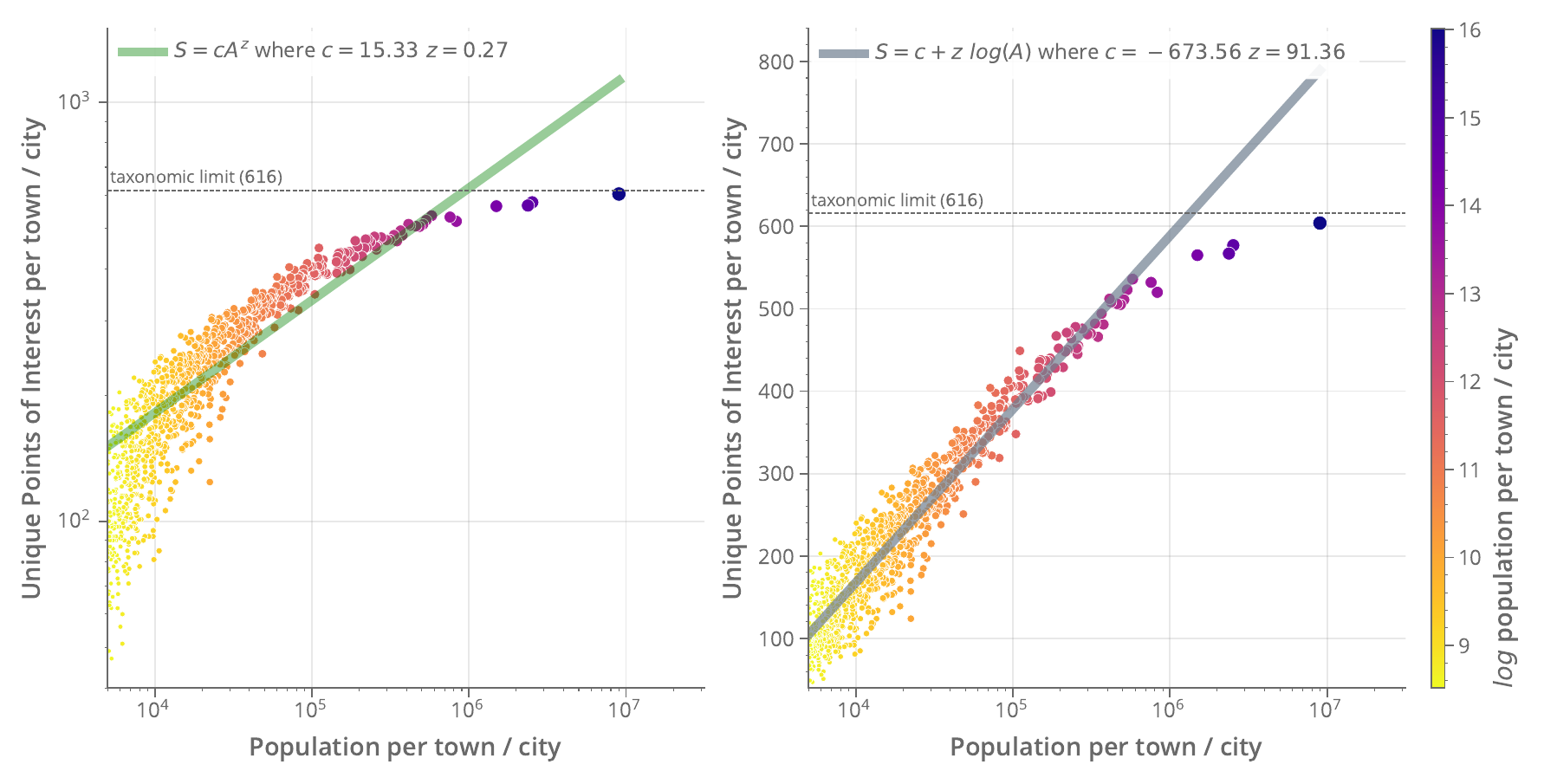}
 \caption[Unique Points Of Interest in relation to population sizes.]{Scatter plots of unique Points Of Interest in relation to population sizes for towns and cities in Great Britain.}\label{fig:global_poi_unique_pop}
\end{figure}

When exploring the relationship between the number of unique Points Of Interest (POI) $S$ and population size $A$ for towns and cities within Great Britain (Figure~\ref{fig:global_poi_unique_pop}) it becomes evident that some form of scaling relationship exists, but that a taxonomic limit is encountered for larger cities. The implication is that the POI classification schema --- which in the case of UK \emph{Ordnance Survey} Points of Interest contains $616$ unique classifications --- is not sufficiently disaggregated to capture the full `species richness' of the largest cities. The lack of a sufficiently granular schema is less of an issue for smaller towns, where the `fish and chip shops' classification may be sufficient in contrast to the enormous culinary diversity of London. The above notwithstanding, there is a clear tendency for POI richness to increase with population, and it could be surmised that this may follow a power-law relationship if a sufficiently diverse classification schema were available.

\begin{figure}[htbp]
 \centering
 \includegraphics[width=\textwidth,height=\textheight,keepaspectratio]{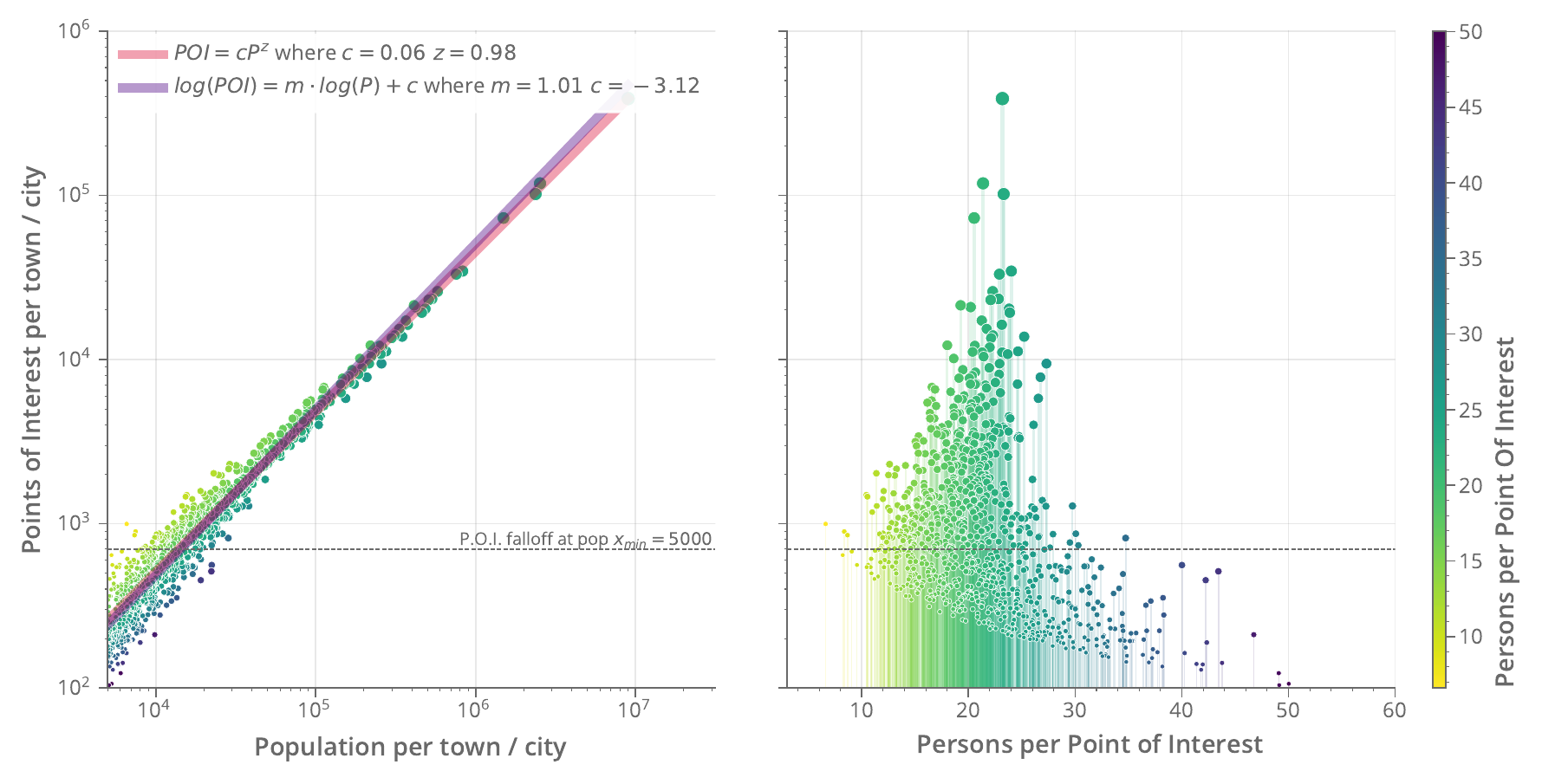}
 \caption[Total Points Of Interest in relation to population sizes]{Scatter plots of the total number of Points of Interest in relation to town and city population sizes for Great Britain (left) and the number of persons per Point Of Interest (right).}\label{fig:global_poi_count_pop}
\end{figure}

When instead looking at the total quantity of POI (Figure~\ref{fig:global_poi_count_pop}) --- as opposed to the unique number thereof --- a linear power-law scaling relationship can be observed against city population. Smaller towns see a higher population variation per POI density, which may be attributable to some locations serving primarily as bedroom communities in contrast to those serving as commercial and administrative hubs for wider districts. As populations increase, the variation narrows until it converges on $\approx 23$ people per POI in London.

The above serves to underscore that the exact quantification of mixed-uses and land-use accessibilities is affected by the resolution and structure of the classification schema. For this reason, mixed-uses cannot necessarily be compared directly between varying schemas unless these have been reduced to an overlapping set of shared classifications.
\section{The quantification of diversity: making sense of a proliferation of indices}\label{the-quantification-of-local-diversity}

The above section discusses the global number of POI for entire towns and cities, and the focus now shifts to how the diversity of land-uses can be measured for a given street-front location within pedestrian walking thresholds.

The extensive application of species diversity measures within ecology provides a rich source of inspiration for their analogous application to mixed land-uses. However, the historical development and application of these methods is convoluted and yields a plethora of indices that muddy rather than clarify the concept of diversity. Rao's quadratic entropy 
\begin{equation}\label{eq:rao_q}
 Rao\ Q = \sum_{i}^{S} \sum_{j \neq{i}}^{S} d_{ij} p_{i} p_{j}
\end{equation}
provides a case-in-point: it sums pairwise species abundances $S$ weighted by species proportions $p_{i}$ and the pairwise dissimilarities $d_{ij}$ between a given species $p_{i}$ and all other species $p_{j}$ \parencite{Rao1982, Rao1982a, Chen2018}. This measure reappears in new guise not just once \parencite{RM1995}, but twice \parencite{Stirling2007}; is related to other indices in mathematically arcane ways \parencite{Ricotta2006, Shimatani2001}; can behave unpredictably depending on the weightings applied to the dissimilarity matrix \parencite{Pavoine2005, Clarke1998, Rogers1999, Clarke1999}; and its partitioning into $\alpha$ (within community), $\beta$ (between community), and $\gamma$ (regional) diversities (which are used by ecologists to describe diversity across different scales) has been described as mathematically \emph{``labyrinthian''} with sometimes unintended or \emph{``meaningless''} outcomes \parencite{MacArthur1965, DeBello2010, Tuomisto2010a}.

Simpler indices exist, including the classic Gini-Simpson diversity index \parencite{Simpson1949} and Shannon information entropy \parencite{Shannon1948}. However, there is a more fundamental issue at work: many of these measures do not behave in accordance with an intuitive conception of diversity because they are quantifying uncertainty or probabilities instead of true species diversity. Further, certain traditional indices such as Rao's quadratic and Gini-Simpson diversity do not behave linearly with the addition of species, thus preventing comparisons between localised collections of species while leading to confounding results, such as where mass species extinctions can have a negligible impact on measured diversity \parencite{Chiu2014, Chao2014}. The solution to this conundrum presents in the form of the Hill diversity index \parencite{Hill1973, Jost2006, Tuomisto2010a} --- a true measure of diversity --- which concisely encompasses concepts broached by traditional methods while transforming these into an elegant form measured in the more interpretable units of equivalent species. Although formalised more than four decades ago, it has only been after the more recent exposition by \textcite{Jost2006} that the Hill index has been widely explored and acknowledged, leading to growing consensus on its use. Critically, the Hill index and its weighted variants adhere to a replication principle, meaning, as with the doubling property, that if two equally sized and equally diverse pools of species were combined, then the diversity would double if there were no overlapping species. It thus becomes feasible to compare the diversity of assemblages across localities meaningfully.

Hill Diversity takes the form
\begin{equation}\label{eq:hill}
 ^{q}D = \Big{(}\sum_{i}^{S}p_{i}^q\Big{)}^{1/(1-q)}\ q\geq0,\ q\neq1
\end{equation}
where probabilities $p_{i}$ are summed for all unique species $S$ while raising the probabilities to the power of the $q$ parameter to control the degree of emphasis on the \emph{richness} of species as opposed to the \emph{balance} of species. In its standard form, the component inside the brackets is known as the \emph{basic sum}, from which the essence of the classic indices can be recovered relative to the strength of $q$ applied to the species probabilities $p_{i}$. The exponent applied to the brackets then maps the basic sum to the effective number of species \parencite{Tuomisto2010a}. As such, $q=0$ causes the formulation to reduce to a simple count of species:
\begin{equation}
 ^{q=0}D = \sum^{S} i\, ,
\end{equation}
which is synonymous with the `species richness' measure
\begin{equation}
 S = \sum^{S} i\, ,
\end{equation}
thereby giving full consideration to all species regardless of how rare or abundant. On the other-hand, $q=2$ places the emphasis on the balance of species
\begin{equation}
 ^{q=2}D = \bigg{(} \sum_{i}^{S} p_{i}^2 \bigg{)}^{-1}
\end{equation}
thereby favouring evenly abundant species to the exclusion of rare species, thus mirroring the intent of Gini-Simpson diversity
\begin{equation}\label{eq:simpson}
 H = 1 - \sum_{i}^{S} p_{i}^2\, .
\end{equation}
$q=1$ offers a middle-ground but is undefined in the limit where $q \to 1$ (because of division through zero), for which purpose it can be found as the exponential of Shannon information entropy
\begin{equation}
 \lim_{q\to1}\ ^{q}D=\exp\Big{(}-\sum_{i}^{S}\ p_{i}\ \log\ p_{i}\Big{)}\, ,
\end{equation}
which, as may be anticipated, is a transformation of Shannon information:
\begin{equation}\label{eq:shannon}
 H = -\sum_{i}^{S}\ p_{i}\ \log\ p_{i}\, .
\end{equation}

It is sometimes beneficial or necessary to take the amount of inter-species disparity or `distance' into account, thus accommodating the degree of difference between species. To this end, the Hill index has more recently been generalised into an extended framework allowing for the computation of weighted forms of diversity \parencite{Chao2014}. In the weighted configuration, probabilities $p_{i}$ are weighted by individual weights $v_{i}$ relative to summed weights $\bar{V}$, where the weights can represent considerations such as branch lengths for species phylogenetic trees or nucleotide distances between DNA sequences:
\begin{equation}\label{eq:hill_generalised}
\begin{aligned}
 ^{q}D(\bar{V}) &= \Bigg{[}\sum_{i}^{S}v_{i}\bigg{(}\frac{p_{i}}{\bar{V}}\bigg{)}^{q} \Bigg{]}^{1/(1-q)}\, ;
 & \bar{V}
 &= \sum_{i}^{S}v_{i}p_{i}\, .
\end{aligned}
\end{equation}
As with the unweighted Hill index, behaviour resembling conventional weighted indices can be recovered from this framework depending on the choice of weights or the strength of $q$. Similar to the unweighted form, the index is found in the limit of $q \to 1$ in the exponential of Shannon information entropy
\begin{equation}
 \lim_{q\to1}\ ^{q}D(\bar{V})=\exp\Bigg{(}-\sum_{i}^{S}\ v_{i}\frac{p_{i}}{\bar{V}}\ \log\ \frac{p_{i}}{\bar{V}}\Bigg{)}\, .
\end{equation}
The units for weighted indices can be expressed in terms of the weights applied to the index, i.e.~effective branch lengths or effective pairwise distances and can be converted to the effective number of species, per
\begin{equation}
 ^{q}D(\bar{V}) = \Bigg{[}\frac{^{q}D(\bar{V})}{\bar{V}}\Bigg{]}^{1/\lambda}\, ,
\end{equation}
where $\lambda=1$ for the phylogenetic form and $\lambda=2$ for the functional form.

This framework can be adapted to address weighted diversity in both the \emph{phylogenetic} form \parencite{Chao2010}
\begin{equation}\label{eq:hill_branch}
\begin{aligned}
 ^{q}D(\bar{T}) = \Bigg{[}\sum_{i}^{S}d_{i}\bigg{(}\frac{p_{i}}{\bar{T}}\bigg{)}^{q} \Bigg{]}^{1/(1-q)}
 & \bar{T} = \sum_{i}^{S}d_{i}p_{i}\, ,
\end{aligned}
\end{equation}
where summed weights $\bar{T}$ are based on branch lengths $d_{i}$ relative to a common source (as may be the case with the branches of a species phylogenetic tree), and the \emph{functional} form \parencite{Chiu2014}
\begin{equation}\label{eq:hill_functional}
\begin{aligned}
 ^{q}D(Q) = \Bigg{[} \sum_{i}^{S} \sum_{j\neq{i}}^{S} d_{ij} \bigg{(} \frac{p_{i} p_{j}}{Q} \bigg{)}^{q} \Bigg{]}^{1/(1-q)}
 & Q = \sum_{i}^{S} \sum_{j\neq{i}}^{S} d_{ij} p_{i} p_{j}\, .
\end{aligned}
\end{equation}
where summed weights $Q$ are expressed as pairwise distances $d_{ij}$ over a distance matrix (as may be the case with nucleotide distances between DNA sequences).
\section{Issues relating to the application of diversity measures to urbanism}\label{the-application-of-diversity-measures-to-urbanism}

The use of diversity measures to quantify mixed-uses at a granular scale presents a series of issues that can undermine the analysis's intent. Firstly, as with ecology, a plethora of confounding indices exist, some of which are highly correlated and can be used interchangeably, whereas others can behave differently depending on the scale of analysis and the number of land-uses \parencite{Song2013}. Nevertheless, the most commonly encountered measures tend to be balance-sensitive indices such as Shannon information entropy \parencite[e.g.][]{Kockelman1997, Christian2011} with a more recent exception provided by \textcite{Yue2017} who argue for the use of Hill Numbers. A subsequent issue is that land-use classifications, unlike species taxonomies, can be surprisingly crude: in some cases encompassing only one or two land-uses and otherwise rarely extending beyond a handful of generic classifications such as residential, commercial, parks, and industrial. Such schemas are not sufficiently disaggregated to fully capture mixed-uses in the sense of rich and varied assortments of land-uses facilitating complex patterns of interaction in thriving neighbourhoods. There is a further particularly problematic issue: the manner in which land-uses are aggregated into larger-scale areal zones or grids removes the measures from the context of the streetscape, and these larger-scale aggregations consequently lose relevance to spatial particularities at the smaller scale and cannot be mapped back to street-fronts in a meaningful sense without invoking the \emph{ecological fallacy}. These methods also lead to the \emph{Modifiable Areal Unit Problem} (MAUP), in which the strength of correlations is affected by changes in the variance of the data as a consequence of varying extents, i.e. larger aggregations tend to boost correlations and may otherwise fluctuate given differing zonal configurations with respect to the underlying prevalences and arrangement of the data \parencite{Fotheringham1991, Song2013}.

The compounding effect of balance-sensitive diversity measures, crude classification schemas, and larger-scale aggregation workflows is that mixed-use measures can behave differently to the intended usage of the concept as intimated by urbanists and urban designers. Together, these approaches result in measures quantifying larger-scale land-use balances as opposed to pedestrian-scale land-use diversity. More simply, markedly different locations may be deemed equivalently mixed-use regardless of how heterogeneously land-uses are dispersed at the pedestrian scale; for example, big-box stores located near a collection of residential enclaves could conceivably be construed as equivalently `mixed-use', or possibly more so, than the high-street of a bustling neighbourhood. This point is underscored by \textcite{Gehrke2017}, who emphasise that conventional mixed-use indices are insensitive to the spatial arrangement or heterogeneity of land-uses, and (within the context of gridded aggregation workflows) argue for the use of methods gauging both the composition (the number or proportion of) and the configuration (the spatial arrangement) of mixed-uses. Another aspect of this problem is that an area consisting of equal portions of, say, industrial and park land-uses would have the same mixed-use score as equal-portions commercial and retail. Such methods are consequently questionable for the purpose of quantifying mixed-uses at a pedestrian scale, and researchers have accordingly observed better results from metrics based on walkable land-use accessibilities and related proxies such as density and connectivity \parencite{Brown2009}. \textcite{Manaugh2013}, in turn, specifically question the assumption --- implicit within species-balance-weighted measures --- that even distributions of undistinguished land-uses should be considered superior to uneven distributions and show that the proximity and complementarity of land-uses provide more substantial outcomes than the use of Shannon information entropy. Some of these issues can be addressed by using the Hill diversity index with lower values of $q$ to emphasise richness rather than balance, especially if combined with more granular representations of land-use data. \textcite{Yue2017} take a significant step in this direction by combining the Hill Index with fine-grained point of interest data and likewise find support for measures emphasising mixed-use richness as opposed to mixed-use balance.

A fundamental issue remains unresolved: the aggregation of land-uses into larger-scale zones or grids foregoes a pedestrian-centric perception of mixed-uses along street fronts. For this reason, the ensuing methodology adopts a `moving-window' method which identifies land-use data points dynamically from each point of analysis in regards to distances measured over the street network and does so while taking the direction of approach into account. As such, distances are known from each point of analysis to each reachable land-use, thereby allowing for diversity measures to be applied in a contextually representative manner while facilitating the use of distance-weighted indices that weight nearby unique land-uses more heavily than distant land-uses. An important implication of spatially precise methods is that, because the observations are contextually anchored, the relationships between different variables --- e.g.~network, centralities, land-use accessibilities, and population densities --- remain coupled and can be meaningfully interpreted at the scale of streets. This approach contrasts that of larger scale aggregations which can induce a disassociation between respective measures.
\section{Exploring diversity indices with the cityseer-api package}

The discussion now proceeds to the exploration and comparison of diversity indices using the \code{cityseer-api} \code{Python} package
\ifpaper \parencite{Simons2021b}.
\else (see Section~\ref{the-cityseer-api-Python-package}).
\fi
The analysis is applied to \emph{Ordnance Survey} \emph{Points of Interest} (POI) data for Greater London, consisting of $388,323$ points with precise Eastings and Northings, with each POI classified according to a schema consisting of $9$ groups (e.g. ‘Commercial’), $52$ categories (e.g. ‘Eating and Drinking’), and $616$ classes (e.g. ‘Fish and Chip shops’).

\begin{figure}[htbp]
 \centering
 \includegraphics[width=\textwidth, keepaspectratio]{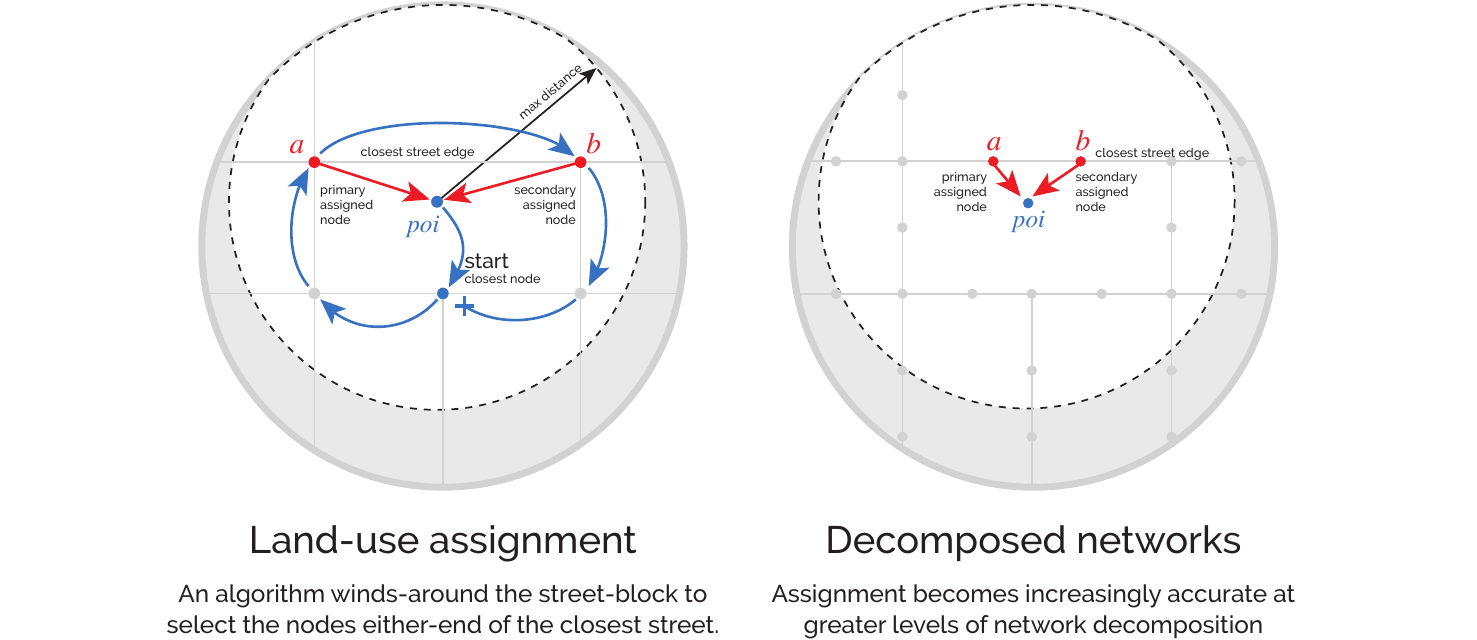}
 \caption[Land-use assignment to the network.]{Land-uses are assigned to the two nodes on either side of the nearest adjacent street edge. An algorithm winds around the street network to encircle the point of interest and identify the closest adjacent street edge. The assignment increases in accuracy when using network decomposition.}\label{fig:poi_assignment_2}
\end{figure}

Points of Interest are assigned to the two street network nodes on either side of the closest adjacent street edge (Figure~\ref{fig:poi_assignment_2}), thus facilitating the dynamic selection of the direction and distance of approach relative to the location of the currently windowed node. Assignment is achieved using a winding algorithm, which first selects the closest adjacent street-network node and then attempts to circle the street network around the point of interest to identify the closest adjacent edge. If encountering a dead-end, the algorithm will backtrack and continue exploring. If exceeding the maximum search distance, it will explore the opposite winding direction to confirm whether any closer edges might exist. Walking-route distances are therefore known from each point of analysis to each of the reachable land-use classifications. It, therefore, becomes feasible to aggregate all land-uses within walking tolerances dynamically and, optionally, to weight their contribution to reflect their proximity to the point of analysis. To further improve the spatial resolution of the analysis, the network used for the remainder of this discussion is `decomposed' so that nodes are located no more than $20m$ apart, which has two important benefits: land-uses can be assigned to the adjacent street network more precisely and the variation of measures can be gauged incrementally along street-lengths.

Building on the discussion in the preceding sections, a series of indices is compared at pedestrian walking thresholds $d_{max}$ ranging from $50m$ to $1600m$:
\begin{itemize}
 \item Traditional indices including Shannon information entropy (Eq:~\ref{eq:shannon}) and Gini-Simpson (Eq:~\ref{eq:simpson});
 \item Hill diversity (Eq:~\ref{eq:hill}) at $q=0$ (emphasises species richness), $q=1$ (intermediate), and $q=2$ (emphasises species balance);
 \item A traditional disparity-weighted index in the form of Rao's quadratic measure (Eq:~\ref{eq:rao_q}), whereby differences between species are weighted more heavily if the amount of disparity is greater. The matrix for class disparities is derived from the \emph{Points of Interest} classification disparities computed in accordance with \textcite{Clarke1999}, resulting in weights of $\frac{3}{3}$, $\frac{2}{3}$, and $\frac{1}{3}$ for large, medium, and small land-use disparities, respectively;
 \item A Hill-based version of disparity-weighted measure (Eq:~\ref{eq:hill_generalised}) using the same pairwise class disparity matrix (as above), and applied at $q=0$, $q=1$, $q=2$.
\end{itemize}

\begin{figure}[htbp]
 \centering
 \includegraphics[width=\textwidth, keepaspectratio]{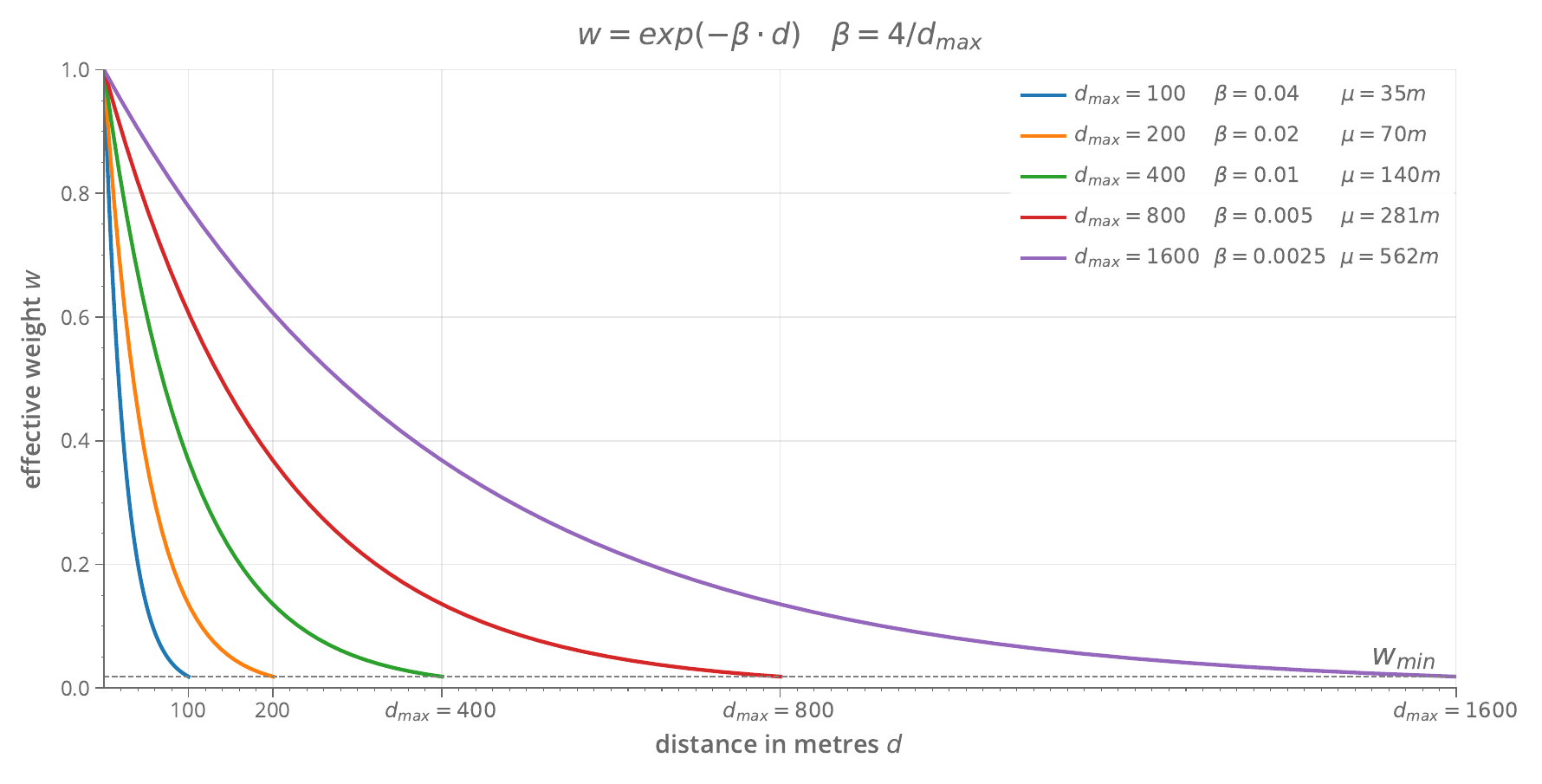}
 \caption[Spatial impedance curves for different $\beta$ parameters.]{Spatial impedance curves for different $\beta$ parameters. Nearer locations can be weighted more heavily than farther locations through use of the negative exponential decay function. The weighted implementations of the Hill Number indices adopt this form of negative exponential to emulate pedestrian spatial impedances, meaning that nearby unique land-uses can be weighted more heavily than distant land-uses, thereby increasing the spatial acuity of the measure in regards to pedestrian walking tolerances.}\label{fig:beta_decays_2}
\end{figure}

Distance-weighted equivalents (where distance is metres) of these measures are also considered. They are derived for a range of spatial impedance curves (see Figure~\ref{fig:beta_decays_2}) where the strength of the $\beta$ parameter reflects a greater or lesser walking tolerance and the decaying curve represents a decreasing willingness to walk correspondingly farther distances. For the proceeding discussion, the $\beta$ are anchored to walking thresholds $d_{max}$ based on a $\beta=4/d_{max}$ conversion, and it is at these points that the threshold $d_{max}$ is enforced so that the curves (and land-use aggregations) don't continue towards infinity. The effective weight $w$ in the $y$ axis reflects the strength of a particular land-use class's contribution to weighted mixed-use measures. Two versions are applied:
\begin{itemize}
 \item A distance-weighted Hill diversity (at $q=0$, $q=1$, $q=2$) in the `phylogenetic' branch-weighted form (Eq:~\ref{eq:hill_branch}), meaning that the contribution of a unique species is weighted by the spatial impedance applied to the shortest path in metres from the currently windowed node of analysis to the closest specimen of the respective land-use class;
 \item Another distance-weighted Hill diversity in the `functional' form (Eq:~\ref{eq:hill_functional}), where walking distances are based on the shortest traversing path connecting the closest specimens for each respective pair of class combinations as routed through the current node.
\end{itemize}
The phylogenetic form of the weighted Hill index assumes the unit of `branch lengths' by definition; however, in this case, the transformation of the distances into effective weights $w$, which scale from 1 at the point of origin towards 0 at the periphery of the selected distance threshold $d_{max}$, already provides a distance-weighted species count without the need for further transformation. This weighting is preferable to the use of the effective species conversions suggested by \textcite{Chao2014} (\ref{eq:hill_generalised}), which would otherwise introduce a form of normalisation which causes complications for comparisons across locations. The distance-weighted functional form is treated similarly, except that the square root is taken so that the squaring implicit in the pairwise formulation is negated.

\begin{figure}[htbp]
 \centering
 \includegraphics[width=\textwidth, height=0.975\textheight, keepaspectratio]{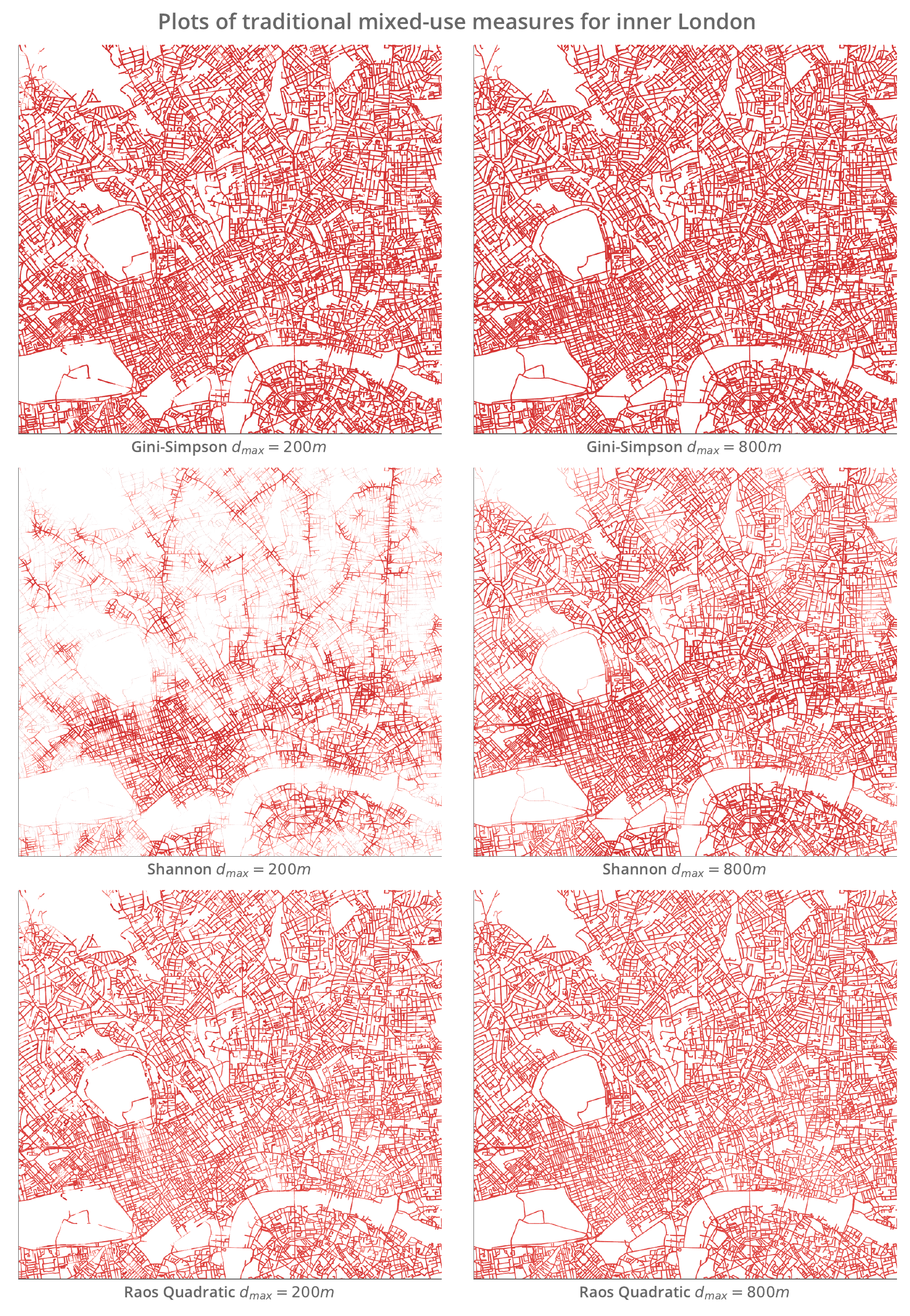}
 \caption[Comparative plots of traditional mixed-use indices for inner London.]{Comparative plots of traditional mixed-use indices (Gini-Simpson, Shannon, and Rao's Quadratic) for inner London at $200m$ and $800m$ network distance thresholds.}\label{fig:diversity_comparisons_traditional}
\end{figure}

\begin{figure}[htbp]
 \centering
 \includegraphics[width=\textwidth, height=0.975\textheight, keepaspectratio]{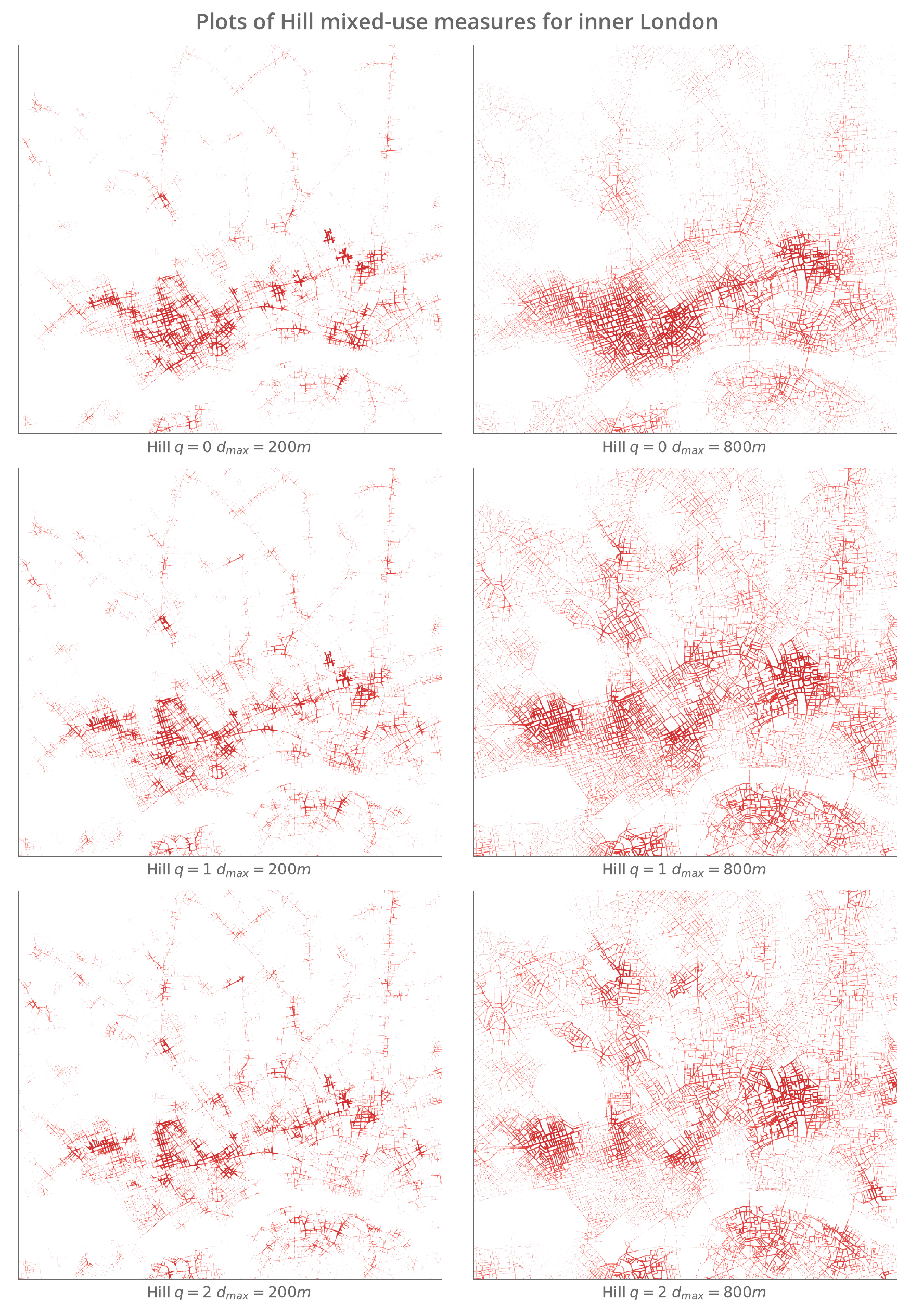}
 \caption[Comparative plots of (unweighted) Hill mixed-use indices for inner London.]{Comparative plots of (unweighted) Hill mixed-use measures at $q=0$, $q=1$, $q=2$ for inner London at $200m$ and $800m$ network distance thresholds.}\label{fig:diversity_comparisons_hill}
\end{figure}

\begin{figure}[htbp]
 \centering
 \includegraphics[width=\textwidth, height=0.975\textheight, keepaspectratio]{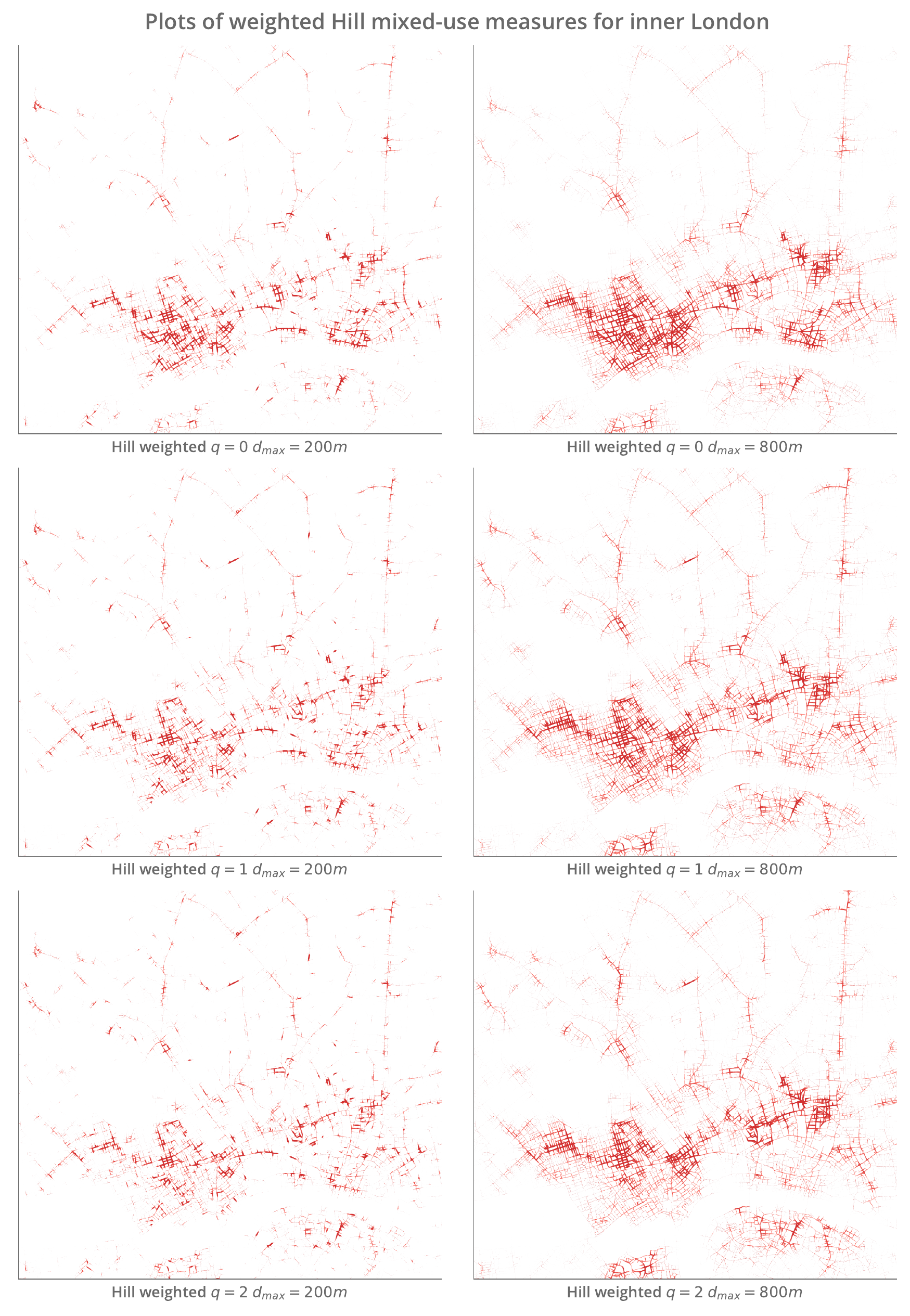}
 \caption[Comparative plots of distance-weighted Hill mixed-use indices for inner London.]{Comparative plots of weighted Hill mixed-use measures at $q=0$, $q=1$, $q=2$ for inner London at $200m$ and $800m$ network distance thresholds.}\label{fig:diversity_comparisons_weighted_hill}
\end{figure}

An intuition for the behaviour of these measures at different distance thresholds $d_{max}=200$ and $d_{max}=800$ is conveyed for the traditional indices in Figure~\ref{fig:diversity_comparisons_traditional}; for unweighted hill indices at $q=0$, $q=1$, $q=2$ in Figure~\ref{fig:diversity_comparisons_hill}; and for the branch-weighted equivalents in Figure~\ref{fig:diversity_comparisons_weighted_hill}. These are further discussed in the next section.

\subsection{Contrasting the behaviour of different indices}\label{comparing-the-behaviour-of-different-indices}

\begin{figure}[htbp]
 \centering
 \includegraphics[width=\textwidth, height=0.975\textheight, keepaspectratio]{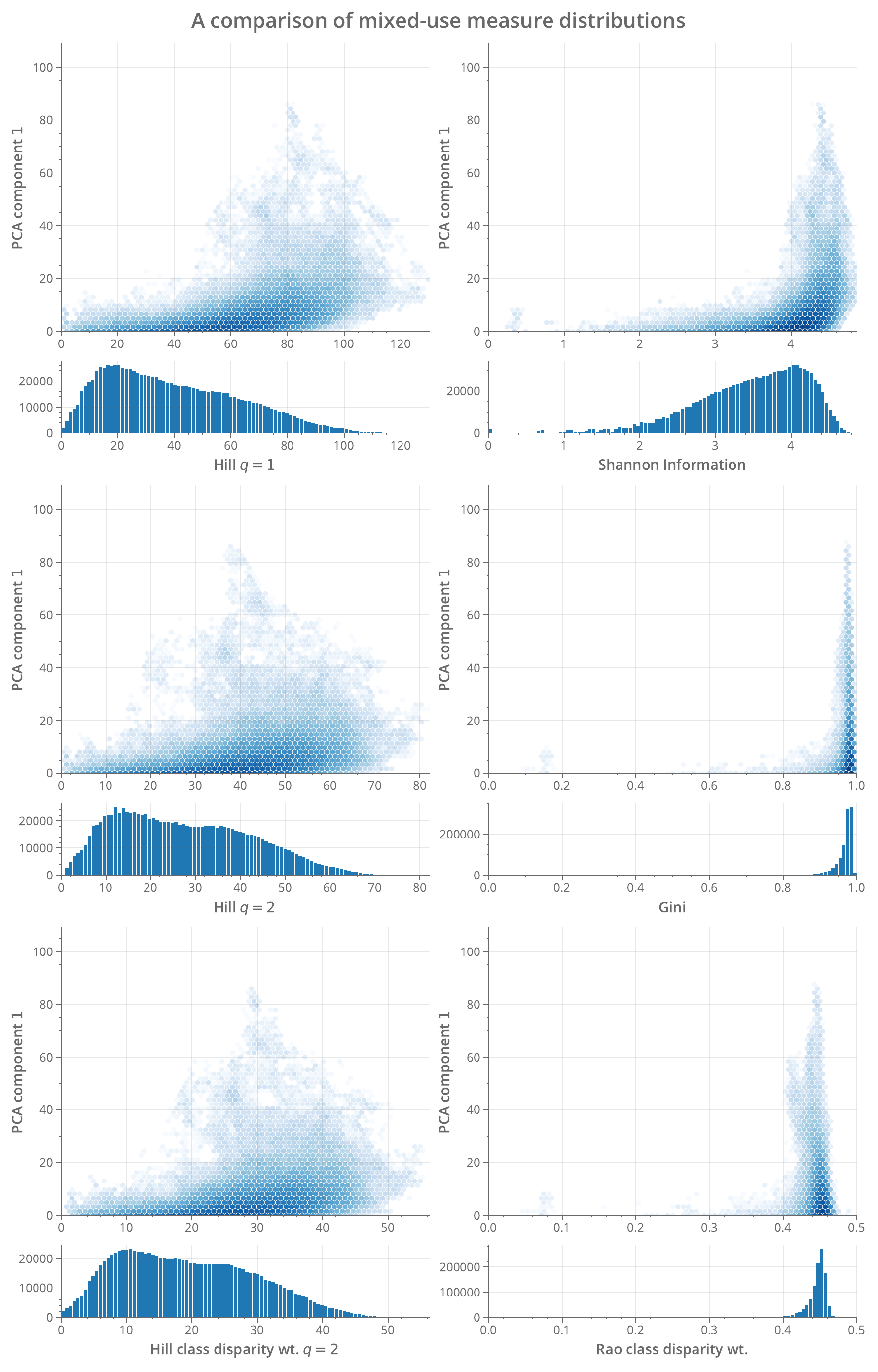}
 \caption[Hexbin and distribution plots of Hill diversity indices compared to non-Hill variants.]{Hexbin and distribution plots of selected Hill diversity indices compared to non-Hill variants of the same type.}\label{fig:mixed_uses_example_distributions}
\end{figure}

\begin{figure}[htbp]
 \centering
 \includegraphics[width=\textwidth, height=0.5\textheight, keepaspectratio]{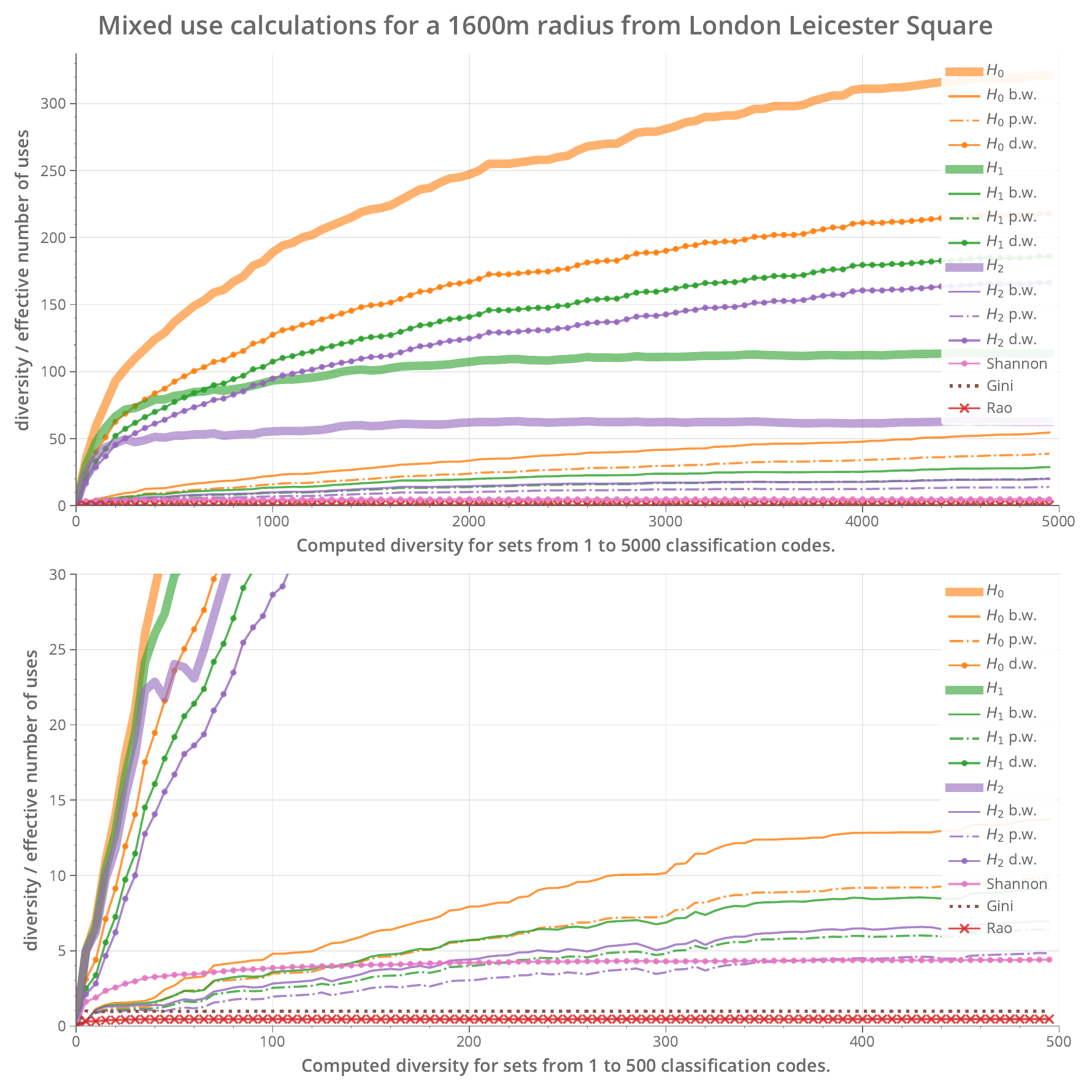}
 \caption[Behaviour of diversity indices for large sets of land-use classifications.]{Behaviour of a range of diversity indices computed for a set of land-use classifications ranging from 1 to 5000 elements (above) using random-removal. Enlargement (below) shows behaviour from 0 to 500 elements. The data is based on the closest 5000 POI land-use classifications as selected from Leicester Square in London. \emph{br.} means `branch', \emph{pw.} means `pairwise', \emph{dist} means `distance', and \emph{wt.} means `weighted'.}\label{fig:mixed_uses_random_removal}
\end{figure}

As explained in Section~\ref{the-quantification-of-local-diversity}, two problematic characteristics of the traditional indices are that they quantify probabilities instead of true diversity and, in the case of Rao and Gini-Simpson diversities, do not necessarily adhere to the replication principle, meaning that differently sized assemblages cannot be compared on a meaningful basis. The implication is that the mathematical behaviour of these indices is not intuitive for the purpose of communicating land-use diversity. Put differently; traditional indices lack `sensitivity' compared to Hill indices of corresponding $q$ values, with the effect particularly noticeable for Gini-Simpson and Rao / Stirling, which are arguably useless from an interpretational point of view, as reflected in the full-page geographic plots for the traditional indices per Figure~\ref{fig:diversity_comparisons_traditional} as compared to the Hill indices per Figure~\ref{fig:diversity_comparisons_hill}. In the comparative distributions and scatterplots of the respective measures in Figure~\ref{fig:mixed_uses_example_distributions}, the traditional indices express the variation of diversity over a compressed numerical range with diminished contrast between areas of higher or lower mixed-uses or between smaller and larger distance thresholds. Whereas it may be possible to recoup some utility from traditional measures through algebraic manipulation (e.g.~logarithmic or boxcox transformations), it should be stressed that the Hill index is already a form of transformation of such indices that offers the distinct advantage of mapping to the effective number of species. Note that whereas Shannon Information is less susceptible to these issues, it still behaves less favourably than its Hill counterpart.

For further emphasis, Figure~\ref{fig:mixed_uses_random_removal} demonstrates how each of the indices behaves as the size of aggregation and the richness of diversity increases. For this purpose, the closest 5000 land-use classifications are selected from Leicester Square in London, a particularly vibrant area with abundant access to diverse assortments of retail, commercial, entertainment, and many other types of land-uses. The respective diversity indices were then repeatedly computed while randomly removing elements (land-use entities) from the set to give a continuous representation of how the different indices scale, with the upper plot showing the diversity measures as computed for a range of 0 to 5000 elements, and the lower plot providing an enlargement from 0 to 500 elements. The Gini-Simpson and Rao / Stirling measures rapidly lose sensitivity. Shannon Information behaves more suitably but lacks the nuance of its Hill transformation. The Hill diversity measures, in contrast, remain intuitive and responsive even for much larger sets of land-uses: for example, $Hill_{q=0}$ indicates approximately 180 unique `species' at 1000 elements, which literally means that there are 180 distinct land-use classifications for this assemblage. When setting $q$ to a higher value, the emphasis shifts from distinct land-use categories to the proportionate balance of species and can still be interpreted in a relatively literal sense of effective species.

Looking at the Hill indices more specifically: the disparity-weighted formulation of Hill diversity emphasises differences between major land-use classifications, such as commercial and education, more heavily than minor differences, such as hospices and hospitals. Its behaviour remains intuitive, though this can be a debatable strategy because even minor differences in land-uses can entail significant interaction, sometimes synergistically so. Similarly, certain major land-use disparities may conversely offer very little potential interaction, as may be the case for manufacturing and parks. The walking-distance-weighted Hill indices increase more linearly throughout the range than the unweighted Hill indices, with the `functional' pairwise form of weighting assuming lower overall values than the `phylogenetic' branch form due to the pairwise end-to-end distances being greater, and therefore weighted less heavily. The benefit of the weighted variants is that they infer greater spatial acuity, which becomes particularly noticeable when applied at larger distance thresholds as seen when comparing the $800m$ plots for the unweighted Hill indices in Figure~\ref{fig:diversity_comparisons_hill} against their weighted implementations in Figure~\ref{fig:diversity_comparisons_weighted_hill}.

\subsection{Land-use accessibility themes as proxies for mixed-uses}\label{land-use-accessibility-themes-as-proxies-for-mixed-uses}

Further exploration is aided through the comparison of correlations for the respective mixed-use measures to two land-use `themes', whereby it is possible to deduce some correspondence between types of mixed-use measures and the distances at which they are applied against two archetypal forms of mixed-uses: 
\begin{itemize}
 \item larger walkable mixed-use districts, and;
 \item smaller locally concentrated mixed-use zones typified by high-streets.
\end{itemize}

\begin{figure}[htbp]
 \centering
 \includegraphics[width=\textwidth, keepaspectratio]{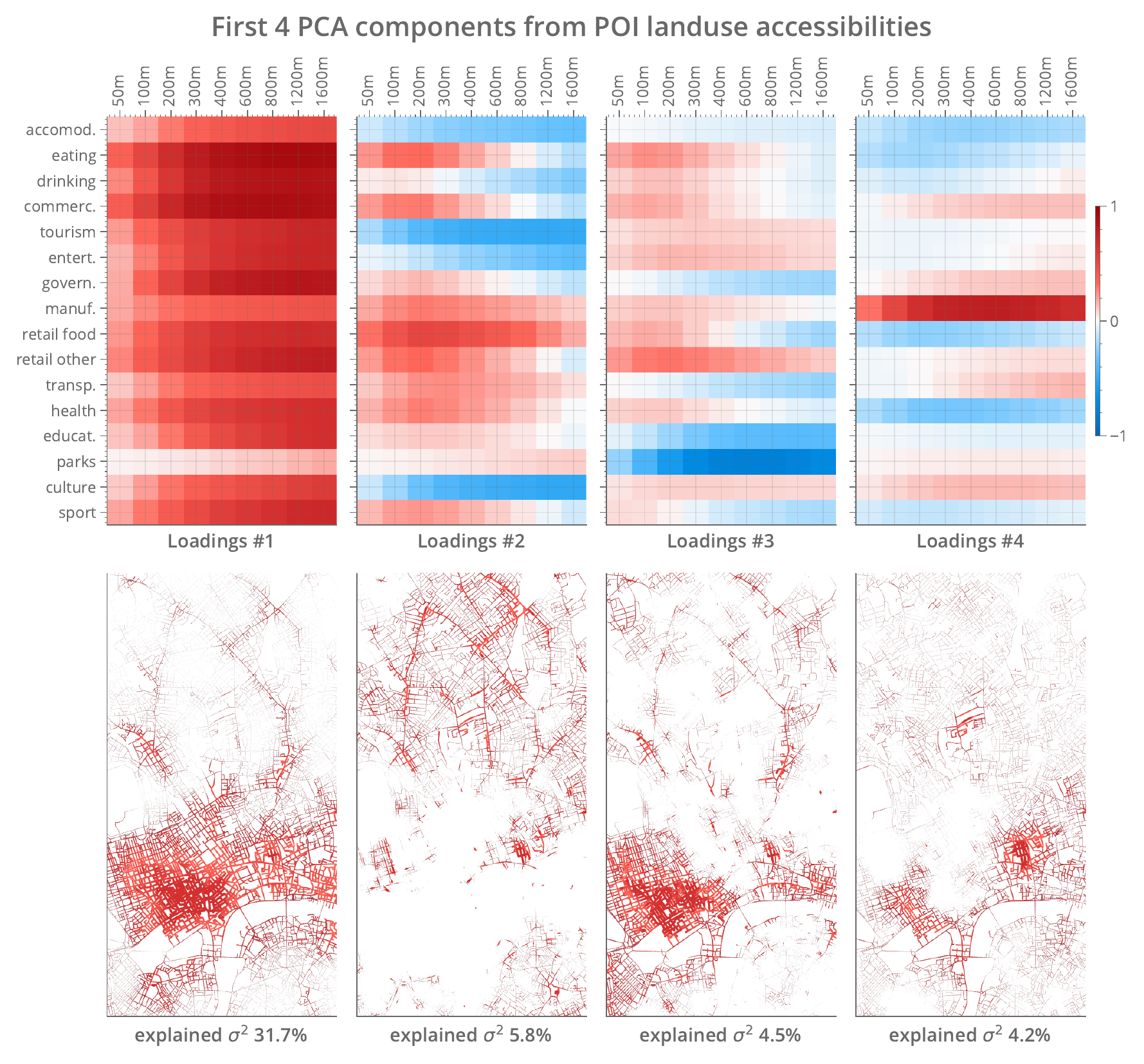}
 \caption[The first four principal components for land-use accessibilities.]{Loadings and geographic plots for the first four principle components for the indicated land-use accessibilities.}\label{fig:pca-a}
\end{figure}

\begin{figure}[htbp]
 \centering
 \includegraphics[width=\textwidth, height=0.95\textheight, keepaspectratio]{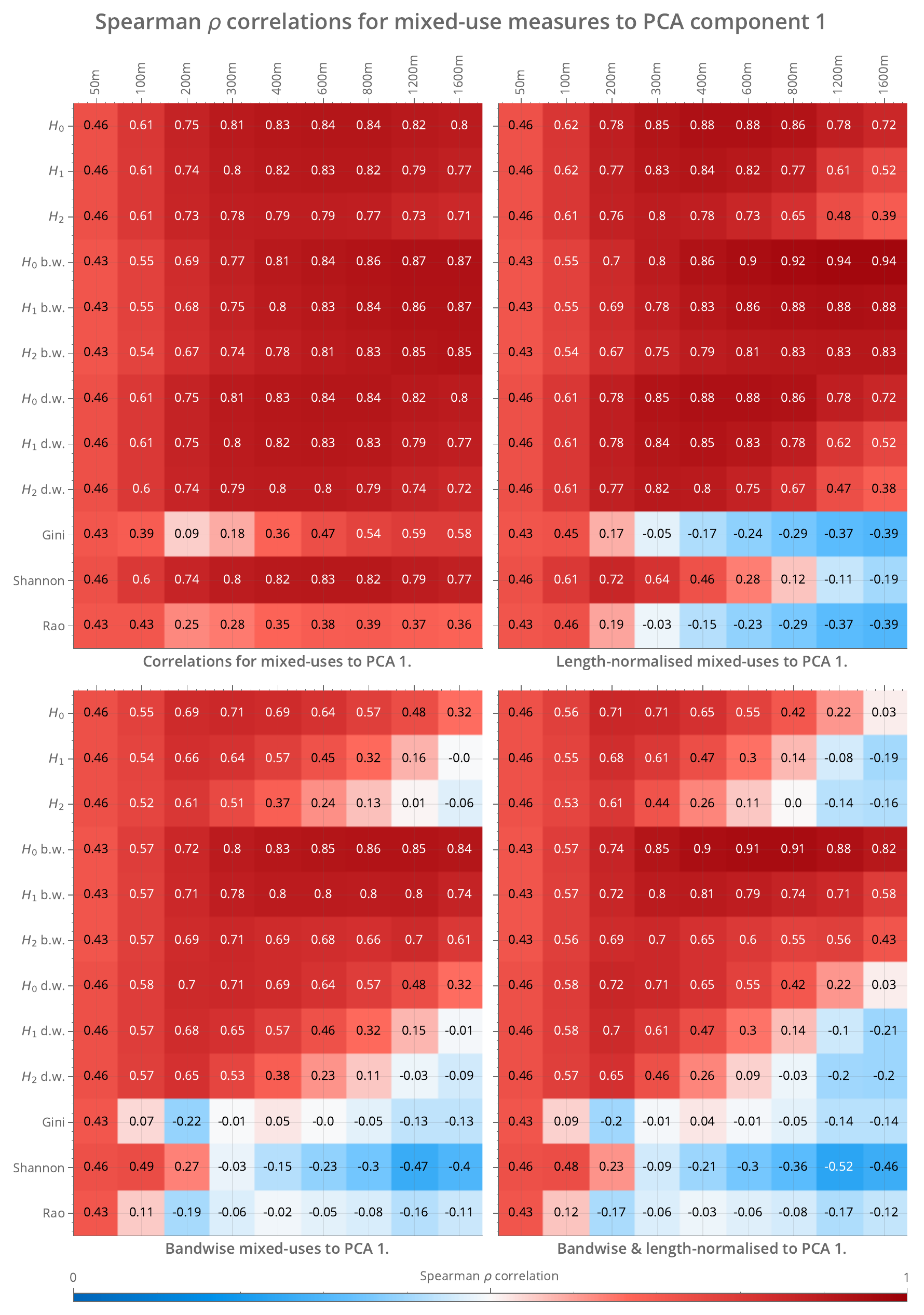}
 \caption[Correlation gridplots for mixed-use indices compared to the first principle component of land-use accessibilities.]{Pearson $\rho$ correlation gridplots for mixed-use indices over a range of distances as compared to the first principle component of land-use accessibilities. Bandwise correlations subtract smaller distance thresholds. Length normalised correlations normalise the number of mixed-uses by street lengths.}\label{fig:mixed_use_measures_correlated_pca_1}
\end{figure}

\begin{figure}[htbp]
 \centering
 \includegraphics[width=\textwidth, height=0.95\textheight, keepaspectratio]{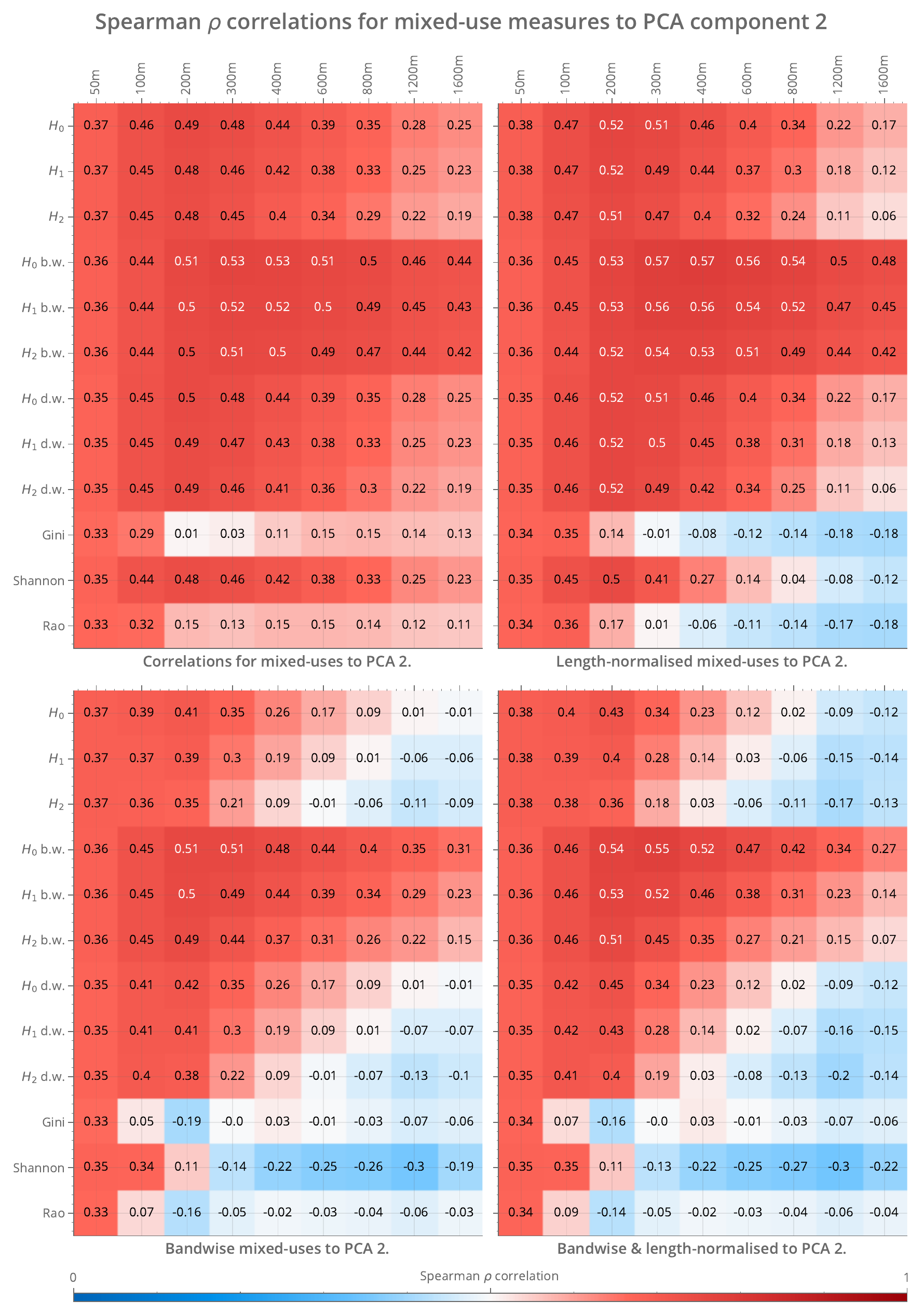}
 \caption[Correlation gridplots for mixed-use indices compared to the second principle component of land-use accessibilities.]{Pearson $\rho$ correlation gridplots for mixed-use indices over a range of distances as compared to the second principle component of land-use accessibilities. Bandwise correlations subtract smaller distance thresholds. Length normalised correlations normalise the number of mixed-uses by street lengths. \emph{H=Hill; b.w.=branch-weighted; d.w.=disparity-weighted.}}\label{fig:mixed_use_measures_correlated_pca_2}
\end{figure}

\begin{figure}[htbp]
 \centering
 \includegraphics[width=\textwidth, height=0.5\textheight, keepaspectratio]{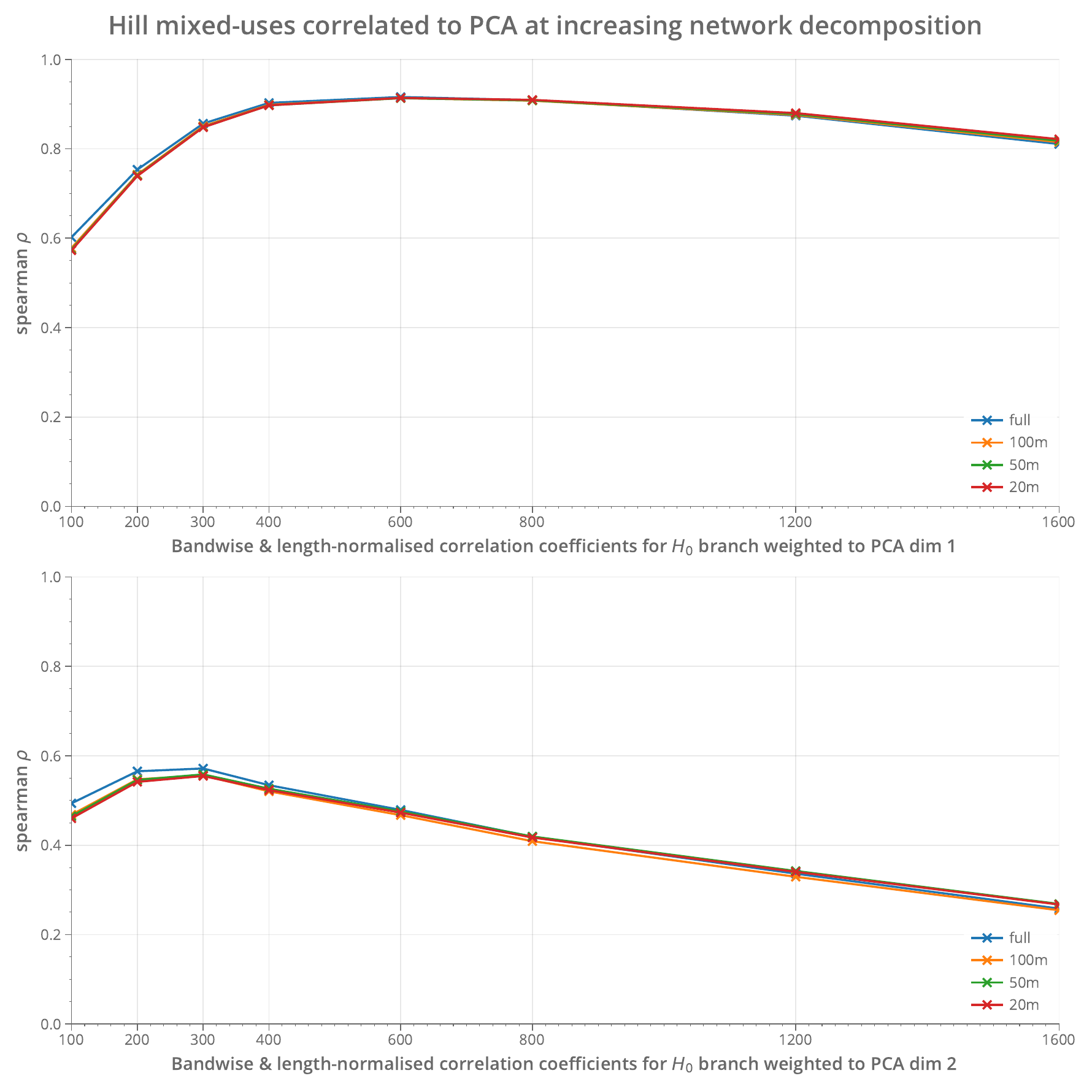}
 \caption[Correlations for Hill mixed-use indices to PCA 1 \& 2 at differing levels of network decomposition.]{Correlations for Hill mixed-use indices to the first two principal components (PCA) of land-use accessibilities, as calculated for different levels of network decomposition. \emph{H=Hill; b.w.=branch-weighted; d.w.=disparity-weighted.}}\label{fig:mixed_use_measures_corr_pca_decompositions}
\end{figure}

The themes are extracted from granular land-use accessibility data also computed with the \code{cityseer-api} package using the same contextual workflows. Principal Component Analysis (PCA), a form of data dimensionality reduction, is then applied to project the raw higher-dimensional data onto a lower-dimensional orthogonal subspace. (The data is mean-centred and scaled to unit variance before proceeding.
\ifpaper A broader discussion of these and other methods is provided in an associated paper, \textcite{Simons2021}.
\else See Section~\ref{untangling-urban-signatures} for a broader discussion of these and other methods.)
\fi
The intuition is that closely correlated or `overlapping' dimensions of data can be `collapsed' into succinct lower-dimensional representations capable of expressing the variance of the original data, thus facilitating the discovery of commonalities or themes in the data which may resonate with intuitive perceptions, otherwise obfuscated by higher-dimensional data spaces.

As shown in Figure~\ref{fig:pca-a}, the land-use accessibility data consists of 16 land-use categories for which distance-weighted accessibilities have been computed at a range of pedestrian walking distances spanning from $100m$ ($\beta=0.04$) to $1600m$ ($\beta=0.0025$). The first four principal components are plotted, showing the spatial manifestation of these dimensions for inner London and the associated \emph{loadings}: effectively, the correlations between the respective Principal Component and the original higher-dimensional features used as inputs. The first component reveals the general intensity of land-uses at the neighbourhood scale and shows high loadings across all land-uses (except parks) and particularly pronounced eating, commercial, government, and retail accessibilities. As such, this component distils patterns of overlap between typically co-located land-uses in areas containing a high intensity and mix of walkable land-uses, drawing out locations such as inner London, Camden Town, Angel, Shoreditch, etc. The second component captures a more locally concentrated mix of land-uses focused on access to eating, commercial, retail, and transportation, thereby proxying local high-streets such as Kentish Town or Stoke Newington and areas with particularly granular or intensive local concentrations of land-uses within inner-London. Note that the first and second components are not mutually exclusive and that locations such as Camden Town may feature prominently in both. Conversely, streets that may feature strongly in the first component will not necessarily have access to a sufficiently locally intensive concentration of land-uses in the same sense as the second component. The first two components together express 37.5\% of the explained variance, with the remainder expressed by successively smaller components that become increasingly specific to particular land-uses or combinations thereof, gradually losing relevance to broader interpretation. The third and fourth components are included for the sake of illustration but are not further considered as part of this analysis.

Figures~\ref{fig:mixed_use_measures_correlated_pca_1} and~\ref{fig:mixed_use_measures_correlated_pca_2} show sets of correlation grid-plots for the respective mixed-use measures as compared to the above-described first two principal components. Each set of grid plots is split into four sections to aid comparison:
\begin{itemize}
 \item The Spearman $\rho$ correlations between the mixed-uses `as measured' and the specified Principle component;
 \item Mixed-use measures which have first been normalised by street lengths before correlation, meaning that the measured mixed-uses are divided by the corresponding street-lengths over which the measures have been applied, thus yielding a mixed-uses-per-street-length form;
 \item Correlations for `bandwise' mixed-uses where each distance threshold subtracts preceding thresholds. The motivation for these adjustments is to explore correlations by distance band, such that nearer intensities of mixed-uses do not cloud correlations for farther thresholds;
 \item both length and bandwise adjusted.
\end{itemize}

In keeping with \textcite{Yue2017} and \textcite{Manaugh2013}: the $Hill_{q=0}$ species-richness variants show higher correlations than the species-balance sensitive variants, with correlations decreasing as $q$ increases. Overall, weighted measures produce higher correlations than their unweighted counterparts regardless of the principal component, bandwise adjustments, or length-normalised adjustments.

Gini-Simpson and Rao demonstrate the weakest correlations, with Shannon Information Entropy faring reasonably well against its $Hill_{q=1}$ counterpart, but falling behind once normalising by street lengths or isolating by band. The distance-weighted variants of Hill diversity find the strongest correlations in all cases, with the branch-weighted versions ultimately edging out the pairwise-weighted variant (the latter of which is not shown due to the very similar performance of the two measures).

The above analysis uses the $20m$ decomposed network, which is compared to the $full$, $100m$, and $50m$ equivalents for the branch-weighted Hill measures in Figure~\ref{fig:mixed_use_measures_corr_pca_decompositions}. Decomposition has a mostly negligible effect on correlations against the Principal Component Analysis, showing marginal decreases for nearer distances and marginal increases for farther distances. These differences likely stem from changes to the underlying distributions due to more incrementally sampled datasets as the level of decomposition increases. In the same sense as the central limit theorem, this increasing level of resolution does not substantially change the observations from a statistical point of view but does confer the benefit of an increased spatial resolution of analysis.
\section{Summary}

The concept of mixed-uses describes access to diverse assortments of land-uses. Importantly, urbanists are ordinarily referring to this mix of uses within a walkable context. Approaches consisting of crude land-use taxonomies, balance-sensitive diversity indices, and larger-scale aggregations are not suitable for this purpose if the results are to be interpreted with contextual specificity for a given location. Put differently; such approaches do not quantify the degree of mixed-uses from a pedestrian's point of view.

These issues are here addressed through a combination of strategies working to preserve contextual information at the pedestrian scale, broadly encompassing the use of granular Point Of Interest land-use classifications; contextually sensitive assignment of land-uses to adjacent street edges; and dynamic workflows consisting of a moving-window algorithm applied to a decomposed network. Calculations are applied at pedestrian walking tolerances up to $1600m$, with distances computed over the street network from each point of analysis to each reachable land-use.

Principal Component Analysis (PCA) is used to proxy characteristic patterns of land-use accessibilities for Greater London at pedestrian walking tolerances. The first principal component reflects a high intensity and mix of walkable land-uses at the neighbourhood scale, and the second accentuates locally concentrated mixes of land-uses more typical of high-streets. In both cases, Hill indices correlate more strongly than the non-Hill variants; distance-weighted Hill indices improve the spatial precision and strengthen the correlations; Hill species richness variants ($q=0$) show stronger correlations than balance-sensitive equivalents, and network decomposition increases the resolution of analysis without having a detrimental impact on the observed correlations.

When targeting walkable access to neighbourhood-scale mixed-uses (per PCA \#1): weighted Hill indices reach maximal correlations in the vicinity of $1000m$ ($\beta=0.004$), else $\approx 500m$ in the case of unweighted equivalents. For cases targeting particularly high local concentrations of high-street mixed-uses (per PCA \#2): weighted Hill indices showed the strongest correlations in the vicinity of $300m$ ($\beta=0.01\bar{3}$), else $\approx 200m$ (or less) in the unweighted case.

The above observations favour the use of species-richness based Hill indices (e.g.~$Hill_{q=0}$). However, caution should be exercised when dealing with simpler land-use classification schemas because data sources containing only a handful of broad classifications (e.g.~\emph{retail}, \emph{business}, \emph{industrial}, \emph{residential}) and will quickly encounter a hard-limit to the number of unique land-uses. In such cases, species-richness measures are less suitable species-balance based methods (e.g.~$Hill_{q=2}$) may need to be considered. Therefore, the nature of the observations and the applicability of findings must be interpreted in the context of the applied land-use classification schema and results stemming from simpler classification schemas may be misconstrued unless mixed-use measures are interpreted judiciously and consistently.

A further conundrum presents: cases attempting to focus on the mix of residential versus non-residential land-uses will encounter the issues mentioned above even if residential land-uses are disaggregated because the low number of residential types places a constraint on the number of non-residential types so as not to skew the analysis away from the intended aim. For this reason, it can be preferable to gauge residential intensities as a separate measure in the form of a localised population density metric, thereby releasing mixed-use measures to focus on granular land-use schemas suited to the quantification of rich mixes of walkable destinations.

\section{Acknowledgements}
\subsection{PhD}

This paper derives from the author's PhD research at the \emph{Centre for Advanced Spatial Analysis}, \emph{University College London}. The author wishes to acknowledge their PhD supervisors, Dr.~Elsa Arcaute and Prof.~Michael Batty, for their gracious support and feedback throughout the development of this work. The author takes sole responsibility for any oversights or shortcomings contained within this paper.

\subsection{Data}

\begin{flushleft}
The geographical plots and statistical figures in this document have been prepared with use of the following sources of data:\linebreak
\linebreak
\textbf{\emph{Ordnance Survey} \emph{Open Roads}}\linebreak
\emph{Contains OS data © Crown copyright and database right 2021.}\linebreak
\linebreak
\textbf{\emph{Ordnance Survey} \emph{Points of Interest} data}\linebreak
\emph{This material includes data licensed from PointX© Database Right/Copyright 2021.}\linebreak
\emph{Ordnance Survey © Crown Copyright 2021. All rights reserved. Licence number 100034829.}\linebreak
\linebreak
\textbf{\emph{UK Data Service} / \emph{Office for National Statistics} census data}\linebreak
\emph{Contains National Statistics data © Crown copyright and database right 2021.} \linebreak
\emph{Contains OS data © Crown copyright and database right (2021).}\linebreak
\end{flushleft}

\section{Citations}
\printbibliography[heading=none]{}
\end{document}